\documentclass[aps,prl,reprint,superscriptaddress,longbibliography]{revtex4-2}

\usepackage{amsmath,amssymb,mathtools,bm,amsthm}
\usepackage{booktabs}
\usepackage{graphicx}
\usepackage{microtype}
\usepackage{tikz}
\usetikzlibrary{arrows.meta,calc,positioning}

\definecolor{AliceBlue}{HTML}{2F5D8A}
\definecolor{BobOrange}{HTML}{BE682C}
\definecolor{Ink}{HTML}{22262B}
\usepackage{hyperref}
\hypersetup{hidelinks}
\newtheorem{theorem}{Theorem}
\newtheorem{corollary}{Corollary}
\newcommand{\C}{\mathbb{C}}

\newcommand{\Id}{I}
\newcommand{\Tr}{\operatorname{Tr}}

\newcommand{\ket}[1]{\lvert #1\rangle}

\newcommand{\braket}[2]{\langle #1\vert #2\rangle}
\newcommand{\norm}[1]{\lVert #1\rVert}
\newcommand{\prlsection}[1]{\par\textit{#1.}---\,}

\begin{document}

\title{Polynomial Separation between One-Way and Two-Way LOCC in Discrimination of Maximally Entangled States}

\author{Youngrong Lim}
\email{sshaep@gmail.com}
\affiliation{Department of Physics, Chungbuk National University, Cheongju 28644, Korea}

\begin{abstract}
One-way local operations and classical communication (LOCC) can fail
to distinguish four orthogonal maximally entangled states in every
local dimension $d\ge4$. We show that every set of $k$ pairwise
orthogonal maximally entangled states in
$\mathbb C^d\otimes\mathbb C^d$ is perfectly distinguishable by an
$A\to B\to A$ protocol whenever $d\ge Ck^4[\log(2k)]^\nu$, for
universal constants $C,\nu>0$. A single feedback message therefore
raises the number of candidates that can always be distinguished
from a constant to $\Omega\!\left(d^{1/4}/(\log d)^{\nu/4}\right)$,
establishing a polynomial separation between one-way and two-way
LOCC. In our protocol, Alice preserves exact orthogonality while
reducing overlaps between Bob's conditional subspaces. Bob reduces the candidates to
a singleton or an orthogonal pair and refines his outcome so that
Alice can complete the discrimination. As a corollary, nine copies
suffice for perfect two-way LOCC discrimination of any orthogonal
maximally entangled set in all sufficiently large local dimensions.
\end{abstract}

\maketitle

\prlsection{Introduction}
Orthogonal quantum states carry classical information that a global
measurement can recover with certainty. When the system is shared
between distant parties, local operations and classical communication
(LOCC) need not recover that information~\cite{Bennett1999,Walgate2000}.
Maximally entangled states (MESs) isolate this restriction. Every
candidate has the same maximally mixed local state, so the label is
encoded entirely in correlations. Distinguishing such states tests
which correlations local measurements can access, a question also
central to quantum data hiding~\cite{Terhal2001}.

The number of candidates and the direction of communication both
matter~\cite{Owari2008,ChitambarHsieh2014}. Any two orthogonal pure
states can be distinguished by
LOCC~\cite{Walgate2000}, and every MES triple is distinguishable at
local dimension $d=3$~\cite{Nathanson2005,Nathanson2013}. In contrast,
one-way LOCC can fail for four MESs in every dimension
$d\ge4$~\cite{Wang2016}. The number of candidates that can always be distinguished by
one-way LOCC therefore remains bounded as the dimension grows. How
does this number scale with local dimension when communication in
both directions is allowed?

Known feedback protocols distinguish specially constructed
ensembles~\cite{Nathanson2013,Tian2015,TianSciRep2016}, and many
criteria exploit the structure of generalized Bell
sets~\cite{Fan2004,Ghosh2004,Tian2016,Wang2019,Li2022,Wang2025}.
These results do not determine how the number of arbitrary MESs
that can always be distinguished scales with dimension. Measurements with
positive partial transpose (PPT) provide a different benchmark. Every
set of at most $d/2+1$ orthogonal MESs is PPT distinguishable
\cite{Nathanson2013}, but such measurements need not admit an LOCC
implementation.

Here we show that a single feedback message raises the number of
candidates that can always be distinguished from a constant to a
quantity growing polynomially with the local dimension. Our bound
applies to \emph{every} orthogonal MES set, without additional
assumptions on its structure. Applying the same result to repeated
preparations also yields a universal copy bound, independent of both
dimension and ensemble size.

\prlsection{Main result}
Every orthogonal MES set in ${\mathbb C}^d\otimes {\mathbb C}^d$ can be written as
\begin{equation}
 \ket{\Psi_i}=(\Id\otimes U_i)\ket{\Phi_d},\qquad
 \Tr(U_i^\dagger U_j)=d\,\delta_{ij},
 \label{eq:mes-representation}
\end{equation}
where each $U_i$ is a unitary acting on Bob's subsystem, $\ket{\Phi_d}=d^{-1/2}\sum_{x=1}^d\ket{xx}$, and $i=1,\ldots,k$. The set is known to both parties, but its prepared label is unknown. No additional shared entanglement is supplied. Our main result is the following.

\begin{theorem}[Universal two-way LOCC discrimination]
\label{thm:main}
    There are universal constants $C,\nu>0$ such that every set in Eq.~\eqref{eq:mes-representation} is perfectly distinguishable by two-way LOCC whenever, for $k\ge3$,
\begin{equation}
 d\ge Ck^4[\log(2k)]^\nu.
 \label{eq:main-bound}
\end{equation}
The protocol consists of one message from Alice to Bob followed by a
single feedback message from Bob to Alice, after which Alice's local
measurement completes the discrimination (pattern $A\to B\to A$). The case $k=2$ follows directly from the Walgate protocol \cite{Walgate2000}.
\end{theorem}

\definecolor{FigBlue}{RGB}{43,92,142}
\definecolor{FigRed}{RGB}{152,46,38}
\definecolor{FigInk}{RGB}{38,40,43}

\begin{figure*}[t]
\centering
\includegraphics[width=0.985\textwidth]{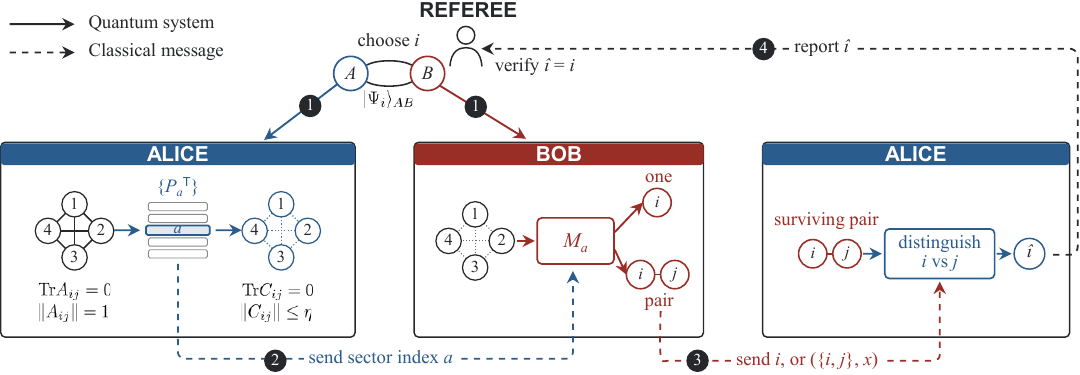}
\caption{
Two-way discrimination protocol, drawn for \(k=4\). A referee chooses
\(i\), prepares the maximally entangled state \(\ket{\Psi_i}_{AB}\),
and distributes the two subsystems. Graph nodes are the candidate
labels and edges their pairwise couplings, initially the relative
unitaries \(A_{ij}=U_i^{\dagger}U_j\). Alice measures the sector
projectors \(\{P_a^{\mathsf T}\}\) of
Eqs.~\eqref{eq:balanced-sectors}--\eqref{eq:eta-q} and announces
\(a\). On the selected sector the couplings compress to \(C_{ij}\)
[Eq.~\eqref{eq:conditional-matrices}], with \(\Tr C_{ij}=0\) (exact
orthogonality) and \(\lVert C_{ij}\rVert\le\eta=1/[4(k-1)]\) (weak
residual coupling). Bob's instrument $M_a$ implements the
POVM with singleton and pair outcomes in Eq.~\eqref{eq:povm}.
For a pair outcome, he also performs the first local step of a
protocol for distinguishing the pair and sends its label together with the refined outcome
$x$ to Alice. Alice then completes the discrimination. Solid arrows follow the quantum systems, and dashed arrows indicate classical messages. The circled numbers give the temporal
order.
}
\label{fig:protocol}
\end{figure*}

The same dimension bound also limits the number of copies needed
for perfect discrimination, when each party can jointly process its
local copies.

\begin{samepage}
\begin{corollary}[A universal copy bound]
\label{cor:copies}
There is a universal integer $N_0$ such that every orthogonal MES
set, in every local dimension $d\ge2$, is perfectly distinguishable
from $N_0$ copies by an $A\to B\to A$ protocol. For all sufficiently
large $d$, nine copies suffice.
\end{corollary}
\end{samepage}

To see this, group Alice's $N$ subsystems and Bob's $N$ subsystems.
The candidates become MESs of local dimension $d^N$,
\begin{align}
 \ket{\Psi_i}^{\otimes N}
 &=\bigl(\Id\otimes U_i^{\otimes N}\bigr)\ket{\Phi_{d^N}},\nonumber\\[-2pt]
 \Tr\!\left[(U_i^\dagger U_j)^{\otimes N}\right]
 &=\left[\Tr(U_i^\dagger U_j)\right]^N=0\quad(i\ne j).
 \label{eq:multicopy}
\end{align}
Since $k\le d^2$, the dimension condition follows from
$d^N\ge Cd^8[\log(2d^2)]^\nu$. For $N=9$ it holds for all
sufficiently large $d$. A universal increase in $N$ covers the
remaining dimensions.

For general orthogonal pure states, the usual LOCC bound is
$k-1$ copies~\cite{Walgate2000,Bandyopadhyay2011}. Two copies suffice
for subsets of the canonical maximally entangled
basis~\cite{Ghosh2004}. Corollary~\ref{cor:copies} removes this basis
restriction while keeping the copy count independent of $k$ and
$d$. The local operations may couple different copies, so the
guarantee is not restricted to measurements that process one copy at a time.

The protocol for a single copy is summarized in Fig.~\ref{fig:protocol}.
Alice's initial measurement gives her no information about the
label. Instead, it changes Bob's identical local states into states
supported on weakly overlapping subspaces, while keeping the joint
candidates exactly orthogonal. Bob can then restrict the possible
labels to a singleton or a pair. His feedback tells Alice which
pair remains and which final measurement resolves it.

\prlsection{Alice's measurement preserving orthogonality}
For any POVM $\{E_a\}$ on Alice's subsystem,
$p(a|i)=\Tr(E_a)/d$ is independent of $i$. Alice's task is therefore to
prepare useful conditional correlations rather than to extract the
label. Alice performs a projective measurement
$\{P_a^{\mathsf T}\}_{a=1}^Q$, resolving only the sector label $a$
and preserving coherence within that sector. Here the $P_a$ are
mutually orthogonal projectors summing to $\Id$.

We write Alice's measurement projectors as $P_a^{\mathsf T}$,
so that their action on the candidates corresponds to
$U_i\mapsto U_iP_a$. With the convention
$\operatorname{vec}(X)=\sum_x\ket{x}\otimes X\ket{x}$, this follows from
$(P_a^{\mathsf T}\otimes\Id)\operatorname{vec}(U_i)
=\operatorname{vec}(U_iP_a)$. For $A_{ij}=U_i^\dagger U_j$ with
$i\ne j$, we require
\begin{equation}
 \Tr(P_aA_{ij})=0,
 \qquad
 \norm{P_aA_{ij}P_a}\le\eta,
 \label{eq:balanced-sectors}
\end{equation}
where
\begin{equation}
 \eta=\frac{1}{4(k-1)},\qquad
 Q\le Ck^2[\log(2k)]^\nu.
 \label{eq:eta-q}
\end{equation}

The two conditions in Eq.~\eqref{eq:balanced-sectors} control
different overlaps. For an outcome of rank $r$, choose an isometry
$J:\C^r\to\C^d$ with $JJ^\dagger=P_a$ and define
\begin{equation}
 W_i=U_iJ,\qquad C_{ij}=W_i^\dagger W_j.
 \label{eq:conditional-matrices}
\end{equation}
The outcome has probability $r/d$ for every candidate. After
identifying Alice's support with $\C^r$ via $J^{\mathsf T}$, the
normalized conditional state is
$\ket{\psi_i}=r^{-1/2}\operatorname{vec}(W_i)$. Hence
\begin{equation}
 \braket{\psi_i}{\psi_j}=r^{-1}\Tr C_{ij}=0,
 \qquad \norm{C_{ij}}\le\eta\quad(i\ne j).
 \label{eq:conditional-properties}
\end{equation}
The trace condition preserves bipartite orthogonality, without which
no later measurement could distinguish the candidates perfectly.
The norm bounds the largest overlap between unit vectors in Bob's
candidate subspaces $\operatorname{ran}W_i$ and
$\operatorname{ran}W_j$. His reduced states, initially all equal to
$\Id/d$, are now maximally mixed on these different rank-$r$
subspaces. Their small overlaps make his next measurement possible.

Constructing sectors that meet both requirements is the main
existence argument. A random decomposition suppresses the operator
compressions~\cite{Meckes2013,Vershynin2018}, but its residual block
traces are generally only small, not zero. We show that small
collective rotations of the projectors cancel every independent
trace constraint while preserving the compression bounds. Refining
the sectors recursively reaches the required suppression without
paying the full dimension cost in a single step. The final sectors
form one projective measurement, so this construction adds no
communication rounds. There are $O(k^2)$ real pairwise
orthogonality constraints, and reducing the subspace overlaps to
$O(1/k)$ contributes a further $O(k^2)$ factor. The resulting
threshold is Eq.~\eqref{eq:main-bound}, up to the stated logarithms.
The full construction and its convergence proof are given in the
Supplemental Material (SM)~\cite{SM}.

\prlsection{Bob's reduction to one or two candidates}
In a one-way protocol from Alice to Bob, Bob must determine the
label. With feedback, he can instead use an instrument whose
outcomes retain at most two candidates while preserving their
purity and orthogonality. Each surviving pair can then be
distinguished by LOCC~\cite{Walgate2000}.

The scalar Gram matrix of the $\ket{\psi_i}$ is already the
identity. For Bob, assemble the conditional isometries as
$\mathcal W=[W_1\ \cdots\ W_k]$ and form their block Gram matrix,
\begin{equation}
 G=\mathcal W^\dagger\mathcal W=[C_{ij}],\qquad C_{ii}=\Id_r.
 \label{eq:gram}
\end{equation}
Since the sum of off-diagonal block norms in every row is at most
$(k-1)\eta=1/4$, this matrix satisfies $G\succeq3\Id/4$.
Its weak off-diagonal blocks can be assigned to positive pair terms,
leaving positive singleton terms on the diagonal,
\begin{equation}
 G=\sum_iX_i+\sum_{i<j}X_{ij}.
 \label{eq:label-graph-decomposition}
\end{equation}
Here $X_i$ is supported only on label $i$, while $X_{ij}$ is
supported on labels $i,j$ and has $(i,j)$ block $C_{ij}$.
This positive decomposition into singleton and pair terms is
established in the SM.

To implement it, put $\mathcal R=\mathcal W G^{-1}$ and
$P_{\mathcal T}=\mathcal W G^{-1}\mathcal W^\dagger$, the projector
onto the span of Bob's candidate subspaces. The effects
\begin{equation}
 \begin{aligned}
 B_{ij}&=\mathcal R X_{ij}\mathcal R^\dagger,\\[-1pt]
 B_i&=\mathcal R X_i\mathcal R^\dagger
       +\frac1k(\Id-P_{\mathcal T})
 \end{aligned}
 \label{eq:povm}
\end{equation}
are positive and sum to the identity. Because
$\mathcal W^\dagger\mathcal R=\Id$, their matrix elements between
candidate vectors reproduce those of $X_{ij}$ and $X_i$.
Thus $B_i$ occurs only for label $i$, and $B_{ij}$ only for labels
$i$ or $j$.

Bob implements a pair outcome with the single Kraus operator
$B_{ij}^{1/2}$. The surviving states remain pure, and their overlap
obeys
\begin{equation}
 \braket{\chi_i}{\chi_j}\propto
 \Tr(W_i^\dagger B_{ij}W_j)=\Tr C_{ij}=0.
 \label{eq:pair-orthogonality}
\end{equation}
Bob then performs the first local step of a Walgate
protocol~\cite{Walgate2000} in which he measures first. He sends
the pair label together with his measurement outcome $x$, and
Alice distinguishes her two orthogonal conditional states. For a
singleton, Bob simply sends
$i$. Alice thus identifies the label with certainty, completing the $A\to B\to A$ protocol and proving Theorem~\ref{thm:main}.

\prlsection{Worst-case discrimination thresholds}
For $d\ge2$, let $m_{\mathsf M}(d)$ be the smallest
size of an orthogonal MES set in $\C^d\otimes\C^d$ that is not
perfectly distinguishable by measurement class $\mathsf M$.
Thus every set with fewer than $m_{\mathsf M}(d)$ candidates is
distinguishable, while at least one set of size $m_{\mathsf M}(d)$
is not. We write $m_{\rightarrow}(d)$ for one-way LOCC, with
communication only from Alice to Bob, and $m_{\leftrightarrow}(d)$
for two-way LOCC, allowing any finite number of communication
rounds.

Inverting the sufficient condition in Theorem~\ref{thm:main} gives
\begin{equation}
 m_{\leftrightarrow}(d)
 =\Omega\!\left[\frac{d^{1/4}}{(\log d)^{\nu/4}}\right].
 \label{eq:cardinality-lower-bound}
\end{equation}
The protocol establishing this lower bound uses only one message
from Alice to Bob and one feedback message. In contrast,
$3\le m_{\rightarrow}(d)\le4$ for every $d\ge4$
\cite{Walgate2000,Wang2016}. The number of candidates that can
always be distinguished therefore grows from a constant under
one-way LOCC to at least the scale in
Eq.~\eqref{eq:cardinality-lower-bound} when feedback is allowed.

For comparison, let $m_{\rm SEP}(d)$ and $m_{\rm PPT}(d)$ denote
the corresponding thresholds for separable (SEP) measurements
and measurements with positive partial transpose (PPT). Every effect of a
separable measurement is a sum of positive product operators,
while every effect of a PPT measurement remains positive under
partial transpose on Bob's subsystem. The inclusions one-way LOCC
$\subseteq$ two-way LOCC $\subseteq$ SEP $\subseteq$ PPT imply
\[
 m_{\rightarrow}(d)\le m_{\leftrightarrow}(d)
 \le m_{\rm SEP}(d)\le m_{\rm PPT}(d).
\]
Consequently, Eq.~\eqref{eq:cardinality-lower-bound} also lower
bounds $m_{\rm SEP}(d)$ and $m_{\rm PPT}(d)$. For PPT it is
superseded by the stronger bound
$m_{\rm PPT}(d)\ge\lfloor d/2\rfloor+2$~\cite{Nathanson2013},
so the new lower bound concerns two-way LOCC and SEP.
Table~\ref{tab:hierarchy} summarizes these thresholds. Further
upper bounds for particular dimensions are given in the SM.

\begin{table*}[t]
\caption{Worst-case thresholds for perfect discrimination of
orthogonal MES sets using one copy.
A lower bound on $m_{\mathsf M}(d)$ guarantees that every smaller
set is distinguishable. An upper bound is witnessed by at least
one indistinguishable set of no more than that size in the
indicated dimensions. The labels $\rightarrow$ and
$\leftrightarrow$ denote one-way LOCC and two-way LOCC with
finitely many rounds, respectively. The two-way lower bound uses only one feedback
message. The notation $d^{1/4-o(1)}$ suppresses the logarithmic
factor in Eq.~\eqref{eq:cardinality-lower-bound}.
$(*)$ denotes this work.}
\label{tab:hierarchy}
\centering
\small
\setlength{\tabcolsep}{9pt}
\renewcommand{\arraystretch}{1.2}
\begin{tabular}{lll}
\toprule
Measurement class $\mathsf M$ & Lower bound on $m_{\mathsf M}(d)$
 & Representative upper bound \\
\midrule
One-way LOCC ($\rightarrow$) & $3$~\cite{Walgate2000}
 & $4$ for all $d\ge4$~\cite{Wang2016} \\
Two-way LOCC ($\leftrightarrow$) & $d^{1/4-o(1)}$ $(*)$
 & $d$ for all $d\ge4$~\cite{Yu2012,YuOh2015} \\
Separable (SEP) & $d^{1/4-o(1)}$ $(*)$
 & $3d/4+1$ for $d\equiv0\pmod4$~\cite{YuOh2015} \\
PPT & $\lfloor d/2\rfloor+2$~\cite{Nathanson2013}
 & $\lfloor7d/8\rfloor+1$ for $d=2^n$, $n\ge4$~\cite{CosentinoRusso2014} \\
\bottomrule
\end{tabular}
\end{table*}

For $d=2^n$ with $n\ge2$, each party holds $n$ qubits. For an
integer $b\ge0$, consider all codebooks of at most $2^b$ orthogonal
MES codewords. Every such codebook is perfectly decodable by class
$\mathsf M$ exactly when $2^b<m_{\mathsf M}(2^n)$.
Equation~\eqref{eq:cardinality-lower-bound} therefore guarantees
two-way decoding for $b$ up to $n/4-O(\log n)$. One-way LOCC
decodes every codebook encoding one bit~\cite{Walgate2000} but
fails on some codebook encoding two bits~\cite{Wang2016}. The
known sets of $d$ states that two-way LOCC cannot distinguish cap
the universally decodable message length at $n-1$
bits~\cite{Yu2012,YuOh2015}. Feedback thus changes
this worst-case message length from one bit to $\Theta(n)$ bits.
This statement concerns every codebook, not the capacity of an
optimized encoding or the number of bits communicated during
decoding.

\prlsection{Beyond maximal entanglement}
The protocol does not require an exactly flat Schmidt spectrum.
Consider orthogonal pure states in $\C^d\otimes\C^D$, $D\ge d$,
with full Schmidt rank $d$ and squared Schmidt coefficients in
$[\lambda_-/d,\lambda_+/d]$, with common bounds
$0<\lambda_-\le1\le\lambda_+$. Put
$\kappa=\lambda_+/\lambda_-$. The same communication pattern
perfectly distinguishes every such set whenever, for $k\ge3$,
\begin{equation}
 d\ge C\kappa^2k^4[\log(2\kappa k)]^\nu.
 \label{eq:bounded-Schmidt}
\end{equation}
Alice's outcomes can now depend on the label, but exact cancellation
of the block traces still preserves orthogonality. The spectral lower bound
ensures positive diagonal terms in Bob's decomposition after a
corresponding reduction of the overlap bound. The case $\kappa=1$
recovers MESs of Schmidt rank $d$ in rectangular systems. The full
argument, including the normalizations for each label, is given in
the SM. This extension concerns orthogonal pure states and does not
assume tolerance to preparation or measurement errors.

\prlsection{Discussion}
One-way LOCC can fail for four MESs, whereas a single feedback
message suffices for every set satisfying the bound relating $d$
and $k$ in Eq.~\eqref{eq:main-bound}. Thus the universally
distinguishable set size grows polynomially with local dimension.
The protocol uses this dimension to preserve orthogonality on every
branch while making Bob's conditional subspaces weakly overlapping,
allowing him to reduce the task to at most two candidates. The same
dimension dependence gives a constant copy bound when each party
jointly processes its local copies.

These statements concern perfect discrimination. For comparison, every MES triple
already admits a one-way protocol with average error at most
$2/(3d)$~\cite{Nathanson2013}. We do not infer a bounded error
separation from the perfect discrimination threshold. Likewise,
the existence of an LOCC protocol with finitely many rounds does
not imply efficient synthesis of exact measurements. Alice's projectors are defined by a
convergent correction procedure, and the universal constants in the
dimension bound have not been optimized or numerically evaluated.

The optimal growth of $m_{\leftrightarrow}(d)$ remains open. A
universal linear threshold would match the scale of the known
upper bounds. Whether further rounds improve the threshold
connects to the role of rounds in other LOCC
tasks~\cite{Chitambar2014,Liu2023}. It is also unknown whether the
LOCC and SEP thresholds coincide,
or how far the universal copy bound can be reduced. Theorem~\ref{thm:main}
settles the triple problem in all sufficiently large dimensions,
but does not determine the remaining cases in smaller dimensions.
Extending the protocol to mixed states or obtaining quantitative
stability under experimental errors would address different
limitations from the exact pure state problem considered here.

The main theoretical results were developed by the author. GPT-6 and Fable-5.1 were used under the author’s direction to assist with exploring and refining mathematical arguments, checking calculations, and revising the manuscript. The author independently verified the final proofs and takes full responsibility for the results and conclusions. This work was supported by the Institute of Information \& Communications Technology Planning \& Evaluation (IITP) grant funded by the Korean government (Ministry of Science and ICT (MSIT)) (No. RS-2024-00437284, No. RS-2025-02283189), and National Research Foundation of Korea (NRF) grant funded by the Korean government (MSIT) (No. RS-2024-00431768).

\bibliography{references}

\end{document}


\title{\texorpdfstring{Supplemental Material for\\}{Supplemental Material for }
``Polynomial Separation between One-Way and Two-Way LOCC in Discrimination of Maximally Entangled States''}

\author{Youngrong Lim}
\email{sshaep@gmail.com}
\affiliation{Department of Physics, Chungbuk National University, Cheongju 28644, Korea}

\maketitle

\noindent
Sections~\ref{sec:proof-main} and~\ref{sec:construction} prove
Theorem~1. The first gives the protocol
assuming Alice's measurement exists, and the second constructs that
measurement. Technical estimates are collected in
Sec.~\ref{sec:technical}. Section~\ref{sec:extensions} gives the
measurement hierarchy, Corollary~1 on multiple copies, and the
extensions to Schmidt spectra and decodable message length.

\section{Overall protocol}
\label{sec:proof-main}

\subsection{Setting and main result}

An orthogonal MES set consists of $k$ states
\begin{equation}
 \ket{\Psi_i}=(\Id\otimes U_i)\ket{\Phi_d},\qquad
 \Tr(U_i^\dagger U_j)=d\,\delta_{ij},\qquad i=1,\dots,k,
 \label{eq:s-mes}
\end{equation}
where $\ket{\Phi_d}=d^{-1/2}\sum_{x=1}^d\ket{xx}$ and the $U_i$ are
unitaries on Bob's system. The set is known to both parties, and
Alice must identify the prepared label with certainty. No additional
shared entanglement is supplied. We restate the main result.

\begin{restatedtheorem}[Universal two-way LOCC discrimination]
There are universal constants $C,\nu>0$ such that every set of the
form \eqref{eq:s-mes} is perfectly distinguishable by two-way LOCC
whenever, for $k\ge3$,
\begin{equation}
 d\ge Ck^4[\log(2k)]^\nu.
 \label{eq:s-main-bound}
\end{equation}
The protocol consists of one message from Alice to Bob followed by a
single feedback message from Bob to Alice, after which Alice's local
measurement completes the discrimination. The case $k=2$ follows
from the Walgate protocol \cite{Walgate2000}.
\end{restatedtheorem}

We write $M_d(\C)$ for the complex $d\times d$ matrices,
$\overline X$ and $X^{\mathsf T}$ for the entrywise complex conjugate
and the transpose in the standard basis, $\norm X$ for the operator
norm, $\hs X=(\Tr X^\dagger X)^{1/2}$ for the Hilbert--Schmidt norm,
and $\RePart A=(A+A^\dagger)/2$ and $\ImPart A=(A-A^\dagger)/(2i)$ for
the Hermitian parts of an operator $A$. For Hermitian $A$ and $B$,
$A\succeq B$ means that $A-B$ is positive semidefinite.

\subsection{Conditional states}
\label{sec:notation}

The basic tool is the transpose trick
$(X\otimes\Id)\ket{\Phi_d}=(\Id\otimes X^{\mathsf T})\ket{\Phi_d}$
for $X\in M_d(\C)$. It gives, for any operator $P$ on Alice's system,
\begin{equation}
 \bra{\Psi_i}(P^{\mathsf T}\otimes\Id)\ket{\Psi_j}
 =\frac1d\Tr(PA_{ij}),\qquad A_{ij}=U_i^\dagger U_j .
 \label{eq:s-vec}
\end{equation}
For $i\ne j$ the relative unitaries $A_{ij}$ are traceless, which is
just the orthogonality of the states.

We write Alice's measurement projectors as $P_a^{\mathsf T}$, so
that their action on the candidates corresponds to $U_i\mapsto U_iP_a$.
Fix an outcome $a$ with $\rank P_a=r$, write $P_a=JJ^\dagger$ for an
isometry $J:\C^r\to\C^d$, and put $W_i=U_iJ$. All quantities below
refer to this outcome. Then
\begin{equation}
 (P_a^{\mathsf T}\otimes\Id)\ket{\Psi_i}
 =\sqrt{\frac{r}{d}}\,
 \bigl(\overline{J}\otimes\Id\bigr)\ket{\psi_i},
 \qquad
 \ket{\psi_i}=(\Id\otimes W_i)\ket{\Phi_r},
 \label{eq:s-conditional-state}
\end{equation}
where $\ket{\Phi_r}=r^{-1/2}\sum_{x=1}^{r}\ket{xx}$. The outcome has
probability $r/d$ for every label, so Alice learns nothing about $i$.
Mapping her support to $\C^r$ with $J^{\mathsf T}$ gives the normalized
state $\ket{\psi_i}=r^{-1/2}\operatorname{vec}(W_i)$, with
$\operatorname{vec}(X)=\sum_x\ket{x}\otimes X\ket{x}$.
Bob's reduced state $W_iW_i^\dagger/r$ is maximally mixed on the
candidate subspace $\mathcal K_i=\operatorname{ran}W_i$, of dimension $r$.

The $r\times r$ matrices $C_{ij}=W_i^\dagger W_j=J^\dagger A_{ij}J$
describe two distinct overlaps,
\begin{align}
 \braket{\psi_i}{\psi_j}
 &=r^{-1}\Tr C_{ij},
 \label{eq:s-overlap-meaning}\\
 \norm{C_{ij}}
 &=\sup_{\norm x=\norm y=1}
   \left|\braket{W_ix}{W_jy}\right|.
 \label{eq:s-subspace-overlap}
\end{align}
The trace gives the overlap of the bipartite states, whereas the
norm is the largest overlap between unit vectors in Bob's candidate
subspaces. Alice must keep the first exactly zero and make the second
small. These are distinct requirements even though the original
states are orthogonal.

\subsection{Alice's measurement}

Alice needs zero block traces and small compressions for every pair
$i<j$. We state the required result for any list of traceless
contractions $A_1,\ldots,A_m$, meaning $\Tr A_\ell=0$ and
$\norm{A_\ell}\le1$. This also covers the extension in
Sec.~\ref{sec:bounded-schmidt}. For MESs, the list consists of the
$m=\binom k2$ relative unitaries $A_{ij}$ with $i<j$. The other pairs
add no conditions because $A_{ji}=A_{ij}^\dagger$.

The number of exact conditions that a single projector $P$ must
satisfy is
\begin{equation}
 s=\dim_{\R}\Span_{\R}\{\RePart A_\ell,\ImPart A_\ell:1\le\ell\le m\}.
 \label{eq:s-spandim}
\end{equation}
Indeed, $\Tr(PA_\ell)=\Tr(P\RePart A_\ell)+i\Tr(P\ImPart A_\ell)$,
with both traces real. Thus the real span counts the independent
linear conditions, and $s\le2m=k(k-1)$ for MESs.

\begin{proposition}[Paving with exact trace balance]
\label{prop:sectors}
For the above traceless contractions, $m\ge1$ and $0<\eta\le1/8$,
there are universal constants $C,\nu>0$ such that
\begin{equation}
 d\ge C(s+1)\eta^{-2}
       \left[\log\!\left(\frac{2m}{\eta}\right)\right]^\nu
 \label{eq:s-final-dimension}
\end{equation}
implies that there is a projective measurement $\{P_a\}_{a=1}^Q$ on
$\C^d$ satisfying
\begin{equation}
 \begin{aligned}
 \Tr(P_aA_\ell)=0,\qquad \norm{P_aA_\ell P_a}\le\eta \quad
     \text{for all }a,\ell, \qquad
 Q\le C\eta^{-2}
       \left[\log\!\left(\frac{2m}{\eta}\right)\right]^\nu.
 \end{aligned}
 \label{eq:s-sector-conclusion}
\end{equation}
\end{proposition}

Paving an operator means splitting the space into blocks on each of
which the compressed operator is small~\cite{Ravichandran2021}, and
trace balance is the extra requirement that every block trace vanish
exactly. The recursive refinement used in the proof only constructs
the final projectors and adds no communication rounds.

For the protocol, apply Proposition~\ref{prop:sectors} to the relative
unitaries $A_{ij}$, $i<j$, with $\eta=1/[4(k-1)]$. For outcome $a$,
Alice applies the Kraus operator $P_a^{\mathsf T}$ and announces $a$.
This measurement preserves coherence within each sector. Since $\Tr(P_aA_{ij})=\Tr C_{ij}$ and
$\norm{P_aA_{ij}P_a}=\norm{C_{ij}}$, every outcome leaves conditional
matrices with
\begin{equation}
 C_{ii}=\Id_r,\qquad \Tr C_{ij}=0,\qquad \norm{C_{ij}}\le\eta
 \qquad(i\ne j).
 \label{eq:s-conditional-data}
\end{equation}

\subsection{Bob's measurement}
\label{sec:bob}

Bob needs positive measurement effects whose outcomes retain at
most two labels. We first decompose the block Gram matrix
$G=[C_{ij}]$ on $\C^k\otimes\C^r$~\cite{HornJohnson2013} into positive
terms supported on single labels or pairs, then realize these terms
as a POVM on Bob's system. They correspond to the nodes and edges in
Fig.~1 of the main text.

\begin{proposition}[Positive decomposition into singleton and pair terms]
\label{lem:gram}
For the data in Eq.~\eqref{eq:s-conditional-data}, $G\succeq\tfrac34\Id$.
For $i\ne j$ put $D_{ij}=(C_{ij}C_{ij}^\dagger)^{1/2}$, so that
$D_{ji}=(C_{ij}^\dagger C_{ij})^{1/2}$ because $C_{ji}=C_{ij}^\dagger$.
Let the pair terms $X_{ij}$ $(i<j)$ and the singleton terms $X_i$ be
the operators on $\C^k\otimes\C^r$ whose only nonzero blocks are
\begin{equation}
 (X_{ij})_{\{i,j\}}
 =\begin{pmatrix}
 D_{ij}&C_{ij}\\
 C_{ij}^\dagger&D_{ji}
 \end{pmatrix},
 \qquad
 (X_i)_{ii}=R_i=\Id_r-\sum_{j\ne i}D_{ij},
 \label{eq:s-edge-blocks}
\end{equation}
where $(X)_{\{i,j\}}$ denotes the $2\times2$ array of blocks of $X$
with row and column labels in $\{i,j\}$. Then
\begin{equation}
 X_{ij}\succeq0,\qquad R_i\succeq\tfrac34\Id_r,\qquad
 G=\sum_iX_i+\sum_{i<j}X_{ij}.
 \label{eq:s-label-graph}
\end{equation}
\end{proposition}

\begin{proof}
For a block vector $x=(x_1,\ldots,x_k)$,
$\langle x,Gx\rangle=\norm x^2+\sum_{i\ne j}\langle x_i,C_{ij}x_j\rangle
\ge\norm x^2-\eta\sum_{i\ne j}\norm{x_i}\norm{x_j}
\ge[1-(k-1)\eta]\norm x^2=\tfrac34\norm x^2$, hence
$G\succeq\tfrac34\Id$. For the pair terms, write the polar
decomposition $C_{ij}=V|C_{ij}|$ with a unitary $V$ and
$|C_{ij}|=(C_{ij}^\dagger C_{ij})^{1/2}=D_{ji}$, so that
$D_{ij}=V|C_{ij}|V^\dagger$. Then $(X_{ij})_{\{i,j\}}=ZZ^\dagger$ with
$Z=\binom{V|C_{ij}|^{1/2}}{|C_{ij}|^{1/2}}$, hence $X_{ij}\succeq0$.
Since $\norm{D_{ij}}=\norm{C_{ij}}\le\eta$, the singleton blocks obey
$R_i\succeq[1-(k-1)\eta]\Id_r=\tfrac34\Id_r$.
In $\sum_iX_i+\sum_{i<j}X_{ij}$, each off-diagonal block receives
$C_{ij}$ from the term supported on $\{i,j\}$, while each diagonal
block is $R_i+\sum_{j\ne i}D_{ij}=\Id_r=C_{ii}$.
Thus the sum equals $G$.
\end{proof}

Let $\mathcal W=[W_1\ \cdots\ W_k]$, so
$\mathcal W^\dagger\mathcal W=G$. Define
\begin{equation}
 \begin{aligned}
 \mathcal R&=\mathcal WG^{-1},&
 P_{\mathcal T}&=\mathcal WG^{-1}\mathcal W^\dagger,\\
 B_{ij}&=\mathcal RX_{ij}\mathcal R^\dagger,&
 B_i&=\mathcal RX_i\mathcal R^\dagger+\frac1k(\Id-P_{\mathcal T}).
 \end{aligned}
 \label{eq:s-povm}
\end{equation}
Here $P_{\mathcal T}$ is the projector onto the span $\mathcal T$
of Bob's candidate subspaces. The columns of $\mathcal R$ form the
dual basis within $\mathcal T$~\cite{Christensen2016}, satisfying
$\mathcal W^\dagger\mathcal R=\Id$. Thus $\mathcal RX\mathcal R^\dagger$
realizes the matrix elements prescribed by $X$ between the columns
of $\mathcal W$.
All effects in Eq.~\eqref{eq:s-povm} are positive, and
Eq.~\eqref{eq:s-label-graph} together with
$\mathcal RG\mathcal R^\dagger=\mathcal WG^{-1}\mathcal W^\dagger=P_{\mathcal T}$
gives
\begin{equation}
 \sum_iB_i+\sum_{i<j}B_{ij}
 =\mathcal RG\mathcal R^\dagger+\Id-P_{\mathcal T}=\Id.
 \label{eq:s-completeness}
\end{equation}
The identities $\mathcal W^\dagger\mathcal R=\Id$ and
$(\Id-P_{\mathcal T})\mathcal W=0$ give
\begin{equation}
 \mathcal W^\dagger B_{ij}\mathcal W=X_{ij},\qquad
 \mathcal W^\dagger B_i\mathcal W=X_i .
 \label{eq:s-support}
\end{equation}
An effect $B$ occurs with probability
$p(B|i')=r^{-1}\Tr(W_{i'}^\dagger BW_{i'})$. By
Eq.~\eqref{eq:s-support}, this vanishes for $B_{ij}$ unless
$i'\in\{i,j\}$, and for $B_i$ unless $i'=i$.

\subsection{Completing the discrimination}
\label{sec:completion}

For a pair outcome, Bob implements $B_{ij}$ with the single Kraus
operator $B_{ij}^{1/2}$. The two possible postmeasurement states then
stay pure, and their overlap is proportional to the trace of
$W_i^\dagger B_{ij}W_j$, which by Eq.~\eqref{eq:s-support} is the
$(i,j)$ block $C_{ij}$ of $X_{ij}$,
\begin{equation}
 \frac1r\Tr(W_i^\dagger B_{ij}W_j)
 =\frac1r\Tr C_{ij}=0.
 \label{eq:s-pair-orthogonality}
\end{equation}
They are therefore exactly orthogonal. Specifying one Kraus operator
is essential because a general instrument with the same effect need
not preserve purity. A Walgate protocol can start with either
party~\cite{Walgate2000}. With Bob first, there is a basis
$\{\ket{b_x}\}$ such that Alice's two conditional states are
orthogonal for every outcome $x$ whenever both are nonzero.
Bob combines the operations into the instrument
$\{\ket{b_x}\bra{b_x}B_{ij}^{1/2}\}_x$ and sends $(\{i,j\},x)$ to
Alice. She then identifies the label with certainty. For a singleton
outcome, Bob sends $i$. This completes the $A\to B\to A$ protocol.

It remains to count dimensions. Here $m=\binom{k}{2}$,
$s\le k(k-1)$ because each pair contributes two Hermitian parts, and
$\eta^{-2}=16(k-1)^2$. Proposition~\ref{prop:sectors} therefore
supplies Alice's measurement whenever
\begin{equation}
 d\ge Ck^4[\log(2k)]^\nu,
 \qquad Q\le Ck^2[\log(2k)]^\nu.
 \label{eq:s-theorem-threshold}
\end{equation}
One factor $k^2$ counts the independent orthogonality conditions, and
the other comes from suppressing Bob's subspace overlaps to $O(1/k)$.
This proves Theorem~1, given Proposition~\ref{prop:sectors}.

\section{Construction of Alice's measurement}
\label{sec:construction}

We prove Proposition~\ref{prop:sectors} in three stages. A random
sector decomposition gives small compressions. A small common
rotation then cancels the block traces without losing the compression
bound. Recursive refinement reaches the required dependence on $\eta$.
Only the final projectors are used in the protocol. The estimates are
stated below and proved in Sec.~\ref{sec:technical}.

\subsection{One refinement step}
\label{sec:onestep}

The first two stages give the following lemma.

\begin{lemma}[One refinement step]
\label{lem:onestep}
Let $m\ge1$ and $A_1,\ldots,A_m\in M_d(\C)$ satisfy $\Tr A_\ell=0$ and
$\norm{A_\ell}\le1$, and let $s$ be defined by Eq.~\eqref{eq:s-spandim}.
There are universal constants $C_0,C_1>0$ such that, for every
$0<\epsilon<1/4$ and every integer $q$ with
\begin{equation}
 q\ge C_0\epsilon^{-2},
 \qquad
 d\ge C_1\!\left[sq^2\log(2q)+\epsilon^{-2}\log(2mq)\right],
 \label{eq:s-assumptions}
\end{equation}
there exist mutually orthogonal projections $P_1,\ldots,P_q$ on $\C^d$
with $\sum_aP_a=\Id_d$, ranks differing by at most one, and
\begin{equation}
 \Tr(P_aA_\ell)=0,
 \qquad
 \norm{P_aA_\ell P_a}\le\epsilon
 \qquad(1\le a\le q,\ 1\le\ell\le m).
 \label{eq:s-conclusion}
\end{equation}
\end{lemma}

If $s=0$, all $A_\ell$ vanish and any balanced decomposition works.
We therefore assume $s\ge1$ below. Balanced ranks keep every sector
between $d/(2q)$ and $2d/q$ once $d\ge2q$, which follows from
Eq.~\eqref{eq:s-assumptions}. They also ensure that no sector is too
small for the next refinement. The protocol itself does not require
equal ranks.

\subsection{Random measurement}
\label{sec:randomsectors}

Let $U$ be Haar random on $U(d)$, put $\bar r=\lfloor d/q\rfloor$,
let $\bar P_1,\ldots,\bar P_q$ be the coordinate projections onto
consecutive blocks of the standard basis with ranks
$r_a\in\{\bar r,\bar r+1\}$ summing to $d$, and set
\begin{equation}
 P_a=U\bar P_aU^\dagger,\qquad a=1,\ldots,q .
 \label{eq:s-random-projections}
\end{equation}
These projectors are the starting point for the correction, not the
final measurement. A sector measurement uses the projectors
$P_a^{\mathsf T}$ as Kraus operators. Measuring individual vectors
$\overline U\ket x$ and discarding the finer result would instead
destroy coherence within a sector and is not the instrument used here.

We use two facts about Haar random unitaries. First, if
$f:U(d)\to\R$ is
invariant under a global phase, $f(e^{i\varphi}U)=f(U)$, and
$L_f$-Lipschitz, meaning $|f(U)-f(V)|\le L_f\hs{U-V}$, then
\begin{equation}
 \Pr\{|f(U)-\E f|>t\}\le 2e^{-cdt^2/L_f^2}
 \qquad(t>0),
 \label{eq:s-concentration}
\end{equation}
where $c>0$ is universal. Phase invariance makes $f$ have the same
distribution on $U(d)$ and $SU(d)$, so Proposition~2.2 of
Ref.~\cite{Meckes2013}, applied to $f$ and $-f$, gives this bound.
Its fluctuation scale is $L_f/\sqrt d$. Second, the Haar averages on
one and two copies give the moments below~\cite{CollinsSniady2006}.
For $d\ge2$ and fixed $X,Y\in M_d(\C)$ they give
\begin{equation}
 \E\!\left[\Tr(UXU^\dagger Y)\right]=\frac{\Tr X\Tr Y}{d},
 \qquad
 \operatorname{Var}\!\left[\Tr(UXU^\dagger Y)\right]
 =\frac{1}{d^2-1}
 \left(\hs{X}^2-\frac{|\Tr X|^2}{d}\right)
 \left(\hs{Y}^2-\frac{|\Tr Y|^2}{d}\right),
 \label{eq:s-weingarten}
\end{equation}
with the variance of a complex random variable $\xi$ defined as
$\E|\xi-\E\xi|^2$.

A matrix element of $A_\ell$ between two fixed vectors of a random
sector has zero mean and fluctuations of order $d^{-1/2}$. A finite
net of test vectors gives the following uniform bound on the
compressions.

\begin{lemma}[Small compressions]
\label{lem:compressions}
Let $A_1,\ldots,A_m$ be traceless contractions on $\C^d$, and let
$P_a=U\bar P_aU^\dagger$, $a=1,\ldots,q$, be the random projectors of
Eq.~\eqref{eq:s-random-projections}, with $U$ Haar random on $U(d)$.
If $d$, $q$, and $\epsilon$ satisfy Eq.~\eqref{eq:s-assumptions} with
$C_0$ and $C_1$ large enough, then
\begin{equation}
 \Pr\left\{\max_{a,\ell}\norm{P_aA_\ell P_a}>\frac\epsilon2\right\}
 \le\frac18 .
 \label{eq:s-compression-event}
\end{equation}
\end{lemma}

Lemma~\ref{lem:compressions} is proved in Sec.~\ref{sec:technical}.

\subsection{Exact orthogonality by a small rotation}
\label{sec:newton}

The exact conditions are linear in the projectors. Choose Hermitian
$H_1,\ldots,H_s$ spanning the real span in Eq.~\eqref{eq:s-spandim},
normalized as $\Tr(H_\alpha H_{\alpha'})=d\,\delta_{\alpha\alpha'}$.
Each $H_\alpha$ is traceless; no bound independent of $d$ on
$\norm{H_\alpha}$ is assumed. Since every $A_\ell$ is a complex
combination of the $H_\alpha$, the conditions $\Tr(P_aA_\ell)=0$ are
equivalent to $\Gamma(P)=0$, where
\begin{equation}
 \mathcal Y=\Bigl\{y\in\R^{q\times s}:\ \textstyle\sum_ay_{a\alpha}=0
 \ \ \forall\alpha\Bigr\},
 \qquad
 \Gamma(P)=\bigl(\Tr(P_aH_\alpha)\bigr)_{a,\alpha}\in\mathcal Y .
 \label{eq:s-phi}
\end{equation}
The residual array $\Gamma(P)$ lies in $\mathcal Y$ because
$\sum_a\Tr(P_aH_\alpha)=\Tr H_\alpha=0$, so there are $s(q-1)$
independent real constraints.

Alice's remaining freedom is a unitary rotation of her whole
measurement, $P_a\mapsto e^KP_ae^{-K}$, whose generator $K$ is
anti-Hermitian, so that $e^K=e^{-i\Lambda}$ with the Hermitian
$\Lambda=iK$. The
linear response of the residuals to an infinitesimal rotation is
\begin{equation}
 D\Gamma_P(K)_{a\alpha}=\Tr\bigl([K,P_a]H_\alpha\bigr).
 \label{eq:s-derivative}
\end{equation}
Only rotations that mix different sectors contribute, because a
rotation inside a single sector commutes with its projector and
leaves every block trace unchanged. The strength of the response is
measured by the smallest singular value of $D\Gamma_P$,
\begin{equation}
 \sigma(P)=\min_{\substack{y\in\mathcal Y\\ \norm y_2=1}}
 \hs{D\Gamma_P^*y},
 \label{eq:s-sigma}
\end{equation}
where the adjoint $D\Gamma_P^*$ refers to the real Hilbert--Schmidt
inner product $\langle K,K'\rangle=\RePart\Tr(K^\dagger K')$ for
generators and to the Euclidean inner product on $\mathcal Y$. Its
explicit form is derived in Sec.~\ref{sec:technical}. If
$\sigma(P)>0$, the map $D\Gamma_P$ has the right inverse of minimum norm
$T=D\Gamma_P^*(D\Gamma_PD\Gamma_P^*)^{-1}$, of norm $1/\sigma(P)$.
A residual $y\in\mathcal Y$ is then removed to first order by the
rotation with generator $-Ty$, whose Hilbert--Schmidt norm is at most
$\norm y_2/\sigma(P)$.

The next lemma ensures that every independent constraint responds
to rotations with a uniform lower bound.

\begin{lemma}[Uniform response]
\label{lem:adjoint}
Let $A_1,\ldots,A_m$ be traceless contractions on $\C^d$ with span
dimension $s\ge1$, let $P_a=U\bar P_aU^\dagger$, $a=1,\ldots,q$,
with $q\ge2$, be the random projectors of
Eq.~\eqref{eq:s-random-projections}, and let
$\sigma(P)$ be the smallest singular value of
Eq.~\eqref{eq:s-sigma}. If $d\ge Csq^2\log(2q)$ with $C$ large
enough, then
\begin{equation}
 \Pr\left\{\sigma(P)<\frac18\sqrt{\frac dq}\right\}\le\frac18 .
 \label{eq:s-invertibility-event}
\end{equation}
\end{lemma}

Lemma~\ref{lem:adjoint} is proved in Sec.~\ref{sec:technical}.

The initial residual is small, because a random measurement is
already nearly balanced. The twirl~\eqref{eq:s-weingarten} with
$X=\bar P_a$ and $Y=H_\alpha$ gives $\E\Tr(P_aH_\alpha)=0$ and
$\operatorname{Var}\Tr(P_aH_\alpha)=r_a(d-r_a)/(d^2-1)$, hence
\begin{equation}
 \E\hs{\Gamma(P)}^2
 =\sum_{a,\alpha}\frac{r_a(d-r_a)}{d^2-1}
 \le s\cdot\frac{d^2}{d^2-1}\le2s .
 \label{eq:s-imbalance}
\end{equation}
By Markov's inequality, $\hs{\Gamma(P)}\le2\sqrt s$ with probability at
least $1/2$. Together with Lemmas~\ref{lem:compressions}
and~\ref{lem:adjoint}, the three events
\begin{equation}
 \max_{a,\ell}\norm{P_aA_\ell P_a}\le\frac\epsilon2,
 \qquad
 \sigma(P)\ge\frac18\sqrt{\frac dq},
 \qquad
 \hs{\Gamma(P)}\le2\sqrt s
 \label{eq:s-good-event}
\end{equation}
hold simultaneously with probability at least $1/4$, so a
measurement satisfying all of them exists.

The last step turns this approximate solution into an exact one.
Starting from $K_0=0$, Alice's rotation generator is corrected against
the current residual,
\begin{equation}
 K_{t+1}=K_t-T\,\Gamma\bigl(e^{K_t}Pe^{-K_t}\bigr),
 \label{eq:s-iteration}
\end{equation}
always with the same $T$. This is a Newton iteration in which the
linear response is frozen at the starting point~\cite{Deuflhard2004},
and the next lemma states when it converges to an exact solution
nearby.

\begin{lemma}[Exact orthogonality from a small residual]
\label{lem:certificate}
Let $A_1,\ldots,A_m$ be exactly traceless contractions with $s\ge1$,
let $\Gamma$ be the residual map of Eq.~\eqref{eq:s-phi}, and let
$\epsilon>0$.
Let $P=(P_a)_{a=1}^q$ be an exact projective decomposition, $q\ge2$, and set
\begin{equation}
 \beta=\hs{\Gamma(P)},\qquad
 \sigma=\sigma(P)>0,\qquad
 b_{\max}=\max_{a,\ell}\norm{P_aA_\ell P_a}.
 \label{eq:s-certificate-data}
\end{equation}
If
\begin{equation}
 \beta\le\frac{\sigma^2}{20\sqrt d},
 \qquad b_{\max}+\frac{4\beta}{\sigma}\le\epsilon,
 \label{eq:s-certificate-conditions}
\end{equation}
the iteration~\eqref{eq:s-iteration} converges to a rotated decomposition
$\tilde P_a=e^KP_ae^{-K}$, with $\hs K\le2\beta/\sigma$, that satisfies
$\Tr(\tilde P_aA_\ell)=0$ and $\norm{\tilde P_aA_\ell\tilde P_a}\le\epsilon$
for every $a,\ell$.
\end{lemma}

The first condition controls convergence of the trace correction;
the second reserves enough compression margin for the rotation.
For MESs, $\beta$ measures the residuals in an independent basis of
orthogonality constraints, while $b_{\max}$ is Bob's largest subspace
overlap. Rigorous bounds on $\beta$, $\sigma$, and $b_{\max}$ certify a
nearby exact measurement. A small residual in floating point
arithmetic alone is insufficient.
Lemma~\ref{lem:certificate} is proved in Sec.~\ref{sec:technical}.

\begin{proof}[Proof of Lemma~\ref{lem:onestep}]
On the simultaneous event in Eq.~\eqref{eq:s-good-event},
$\beta\le2\sqrt s$, $\sigma\ge\frac18\sqrt{d/q}$ and
$b_{\max}\le\epsilon/2$. The conditions of
Lemma~\ref{lem:certificate} then follow from
\begin{equation}
 d\ge Csq^2,\qquad d\ge Csq\epsilon^{-2}.
 \label{eq:s-smallness}
\end{equation}
Both are implied by Eq.~\eqref{eq:s-assumptions}, since
$q\ge C_0\epsilon^{-2}$. Thus an exact balanced decomposition exists
with every compressed norm at most $\epsilon$.
\end{proof}

\subsection{Recursive refinement}
\label{sec:hierarchy}

A single application of Lemma~\ref{lem:onestep} with $\epsilon=\eta$
costs $d\gtrsim s\eta^{-4}$ up to logarithms, a $k^6$ threshold for
MES inputs. The $k^4$ bound of Proposition~\ref{prop:sectors} comes
from applying the lemma recursively. Inside a sector with isometry
$J_v$ onto its range, the compressed operators $J_v^\dagger A_\ell J_v$
are again traceless, because the trace conditions hold exactly, and
after division by their current norm bound they are again
contractions, so Lemma~\ref{lem:onestep} applies inside it. With $q_h$
subsectors and suppression factor $\epsilon_h$ at level $h$, the
construction has the following form.

\begin{center}
\renewcommand{\arraystretch}{1.18}
\begin{tabular}{@{}l@{\hspace{14pt}}l@{\hspace{14pt}}l@{\hspace{14pt}}l@{}}
\toprule
Stage & Number of sectors & Sector dimension & Overlap bound\\
\midrule
Initial & $1$ & $d$ & $1$\\
After refinement $j$ & $Q_j=\prod_{h\le j}q_h$
 & $\lfloor d/Q_j\rfloor$ or $\lceil d/Q_j\rceil$
 & $\rho_j=\prod_{h\le j}\epsilon_h$\\
Final & $Q_L$ & $\lfloor d/Q_L\rfloor$ or $\lceil d/Q_L\rceil$ & $\eta$\\
\bottomrule
\end{tabular}
\end{center}
Every row is a projective decomposition with exactly zero block
traces, and only the final row is Alice's physical measurement.

The suppression factors must be allocated with care. At level $j$ the
lemma is applied in each of the $Q_{j-1}$ sectors already present,
each of dimension about $d/Q_{j-1}$, so the dimension needed is about
$Q_{j-1}$ times the cost of one refinement $sq_j^2$, with
$q_j\propto\epsilon_j^{-2}$. The choice $\epsilon_j=\eta^{1/2^j}$, in
which every level supplies the square root of the suppression still
missing, makes this cost the same at every level. Indeed,
$Q_{j-1}\propto(\epsilon_1\cdots\epsilon_{j-1})^{-2}=\eta^{-2+2^{2-j}}$
and $q_j^2\propto\eta^{-2^{2-j}}$, so $Q_{j-1}sq_j^2\propto s\eta^{-2}$
independently of $j$. After about $\log_2\log(1/\eta)$ levels the
missing factor is a constant, which the last level supplies, and the
constants accumulated over the levels produce the polylogarithmic
factor in Eq.~\eqref{eq:s-final-dimension}.
Section~\ref{sec:technical} shows that every level succeeds under
condition~\eqref{eq:s-final-dimension},
which proves Proposition~\ref{prop:sectors} with
$\nu=1+\log_2(C_0+1)$. Together with Sec.~\ref{sec:completion}, this
completes the proof of Theorem~1.

The constants can be made explicit by tracking the estimates below,
but $C_0$, $C_1$, and the final prefactor $C$ are not evaluated or
optimized here. Numerical evaluation uses matrix products, linear
solves, and matrix exponentials. A finite iterate need not have zero
trace residual, so this is an existence proof for an exact
measurement, not a proof of efficient exact synthesis.
Lemma~\ref{lem:certificate} certifies a nearby exact decomposition
when rigorous bounds on its inputs are available.

\subsection{Technical proofs}
\label{sec:technical}

Besides the concentration bound~\eqref{eq:s-concentration} and Haar
moments~\eqref{eq:s-weingarten}, we use two standard facts. A
$\delta$-net of a unit sphere is a finite set within distance $\delta$
of every unit vector. In real dimension $n$, it can have at most
$(3/\delta)^n$ points for $0<\delta\le1$~\cite{Vershynin2018}.
Banach's contraction principle states that a map taking a closed ball
into itself and contracting distances by a uniform factor below one
has a unique fixed point, to which its iterates converge.

\begin{proof}[Proof of Lemma~\ref{lem:compressions}]
Fix $a$ and $\ell$, and abbreviate $A=A_\ell$ and $r=r_a$.

\emph{Step 1. One matrix element.} Let $x,y$ be unit vectors in the
range of $\bar P_a$. The twirl~\eqref{eq:s-weingarten} with
$X=\ket y\bra x$ gives $\E\bra{Ux}A\ket{Uy}=\braket xy\Tr A/d=0$. The
real part $g(U)=\RePart\bra{Ux}A\ket{Uy}$ is invariant under a global
phase and $2$-Lipschitz, because
$|g(U)-g(V)|\le\norm{(U-V)x}+\norm{(U-V)y}\le2\hs{U-V}$ for a
contraction $A$, and the same holds for the imaginary part. Applying
Eq.~\eqref{eq:s-concentration} to both parts, and noting that
$|z|>t$ forces $|\RePart z|$ or $|\ImPart z|$ to exceed $t/\sqrt2$,
\begin{equation}
 \Pr\bigl\{|\bra{Ux}A\ket{Uy}|>t\bigr\}\le4e^{-cdt^2/8}.
 \label{eq:s-bilinear}
\end{equation}

\emph{Step 2. Finitely many test vectors.} Let $\mathcal N$ be a
$\tfrac14$-net of the unit sphere of the range of $\bar P_a$, which is
a sphere in a real space of dimension $2r$, so
$|\mathcal N|\le12^{2r}$. Choose unit vectors $x,y$ in this range with
$\norm{P_aAP_a}=|\bra{Ux}A\ket{Uy}|$, and net points $x_0,y_0$ within
distance $\tfrac14$ of them. Then
$|\bra{Ux}A\ket{Uy}-\bra{Ux_0}A\ket{Uy_0}|\le\tfrac12\norm{P_aAP_a}$,
and therefore
\begin{equation}
 \norm{P_aAP_a}\le2\max_{x,y\in\mathcal N}|\bra{Ux}A\ket{Uy}| .
 \label{eq:s-net-comparison}
\end{equation}

\emph{Step 3. Union bound.} By Eqs.~\eqref{eq:s-bilinear}
and~\eqref{eq:s-net-comparison} with $t=\epsilon/4$, and by
$|\mathcal N|^2\le12^{4r}\le e^{20d/q}$ for $r\le2d/q$,
\begin{equation}
 \Pr\Bigl\{\norm{P_aAP_a}>\frac\epsilon2\Bigr\}
 \le4|\mathcal N|^2e^{-cd\epsilon^2/128}
 \le4\exp\!\Bigl(\frac{20d}{q}-\frac{cd\epsilon^2}{128}\Bigr).
 \label{eq:s-single-compression}
\end{equation}
If $C_0\ge5120/c$, the condition $q\ge C_0\epsilon^{-2}$ gives
$20d/q\le cd\epsilon^2/256$. A union bound over the $qm$ pairs
$(a,\ell)$ then bounds the failure probability by
$4qm\,e^{-cd\epsilon^2/256}$, which is at most $\frac18$ as soon as
$d\epsilon^2\ge(256/c)\log(32mq)$. This follows from the second
condition in Eq.~\eqref{eq:s-assumptions}.
\end{proof}

\emph{Response formulas.} The proofs of Lemmas~\ref{lem:adjoint}
and~\ref{lem:certificate} use the following explicit form of the
response. For $y\in\mathcal Y$ put
$Y_a=\sum_\alpha y_{a\alpha}H_\alpha$, so that
$\hs{Y_a}^2=d\norm{y_a}_2^2$, where $y_a\in\R^s$ is the $a$-th row of
$y$. For every generator $K$,
$\langle D\Gamma_P(K),y\rangle=\Tr\bigl(K\sum_a[P_a,Y_a]\bigr)
=\bigl\langle K,\sum_a[Y_a,P_a]\bigr\rangle$, so that
\begin{equation}
 D\Gamma_P^*y=\sum_a[Y_a,P_a].
 \label{eq:s-adjoint-def}
\end{equation}
Inserting the resolution $\Id=\sum_bP_b$ on both sides, the diagonal
blocks cancel and the $(a,b)$ block of $D\Gamma_P^*y$ equals
$P_a(Y_b-Y_a)P_b$ for $a\ne b$. Since distinct blocks are orthogonal
in the Hilbert--Schmidt norm,
\begin{equation}
 \hs{D\Gamma_P^*y}^2=\sum_{a\ne b}\hs{P_a(Y_a-Y_b)P_b}^2 ,
 \label{eq:s-adjoint-norm}
\end{equation}
which is the precise form of the statement that only rotations mixing sectors move the residuals. Two bounds follow from these formulas.
Splitting each block into its two terms and using
$\sum_b\hs{P_aY_aP_b}^2=\hs{P_aY_a}^2\le\hs{Y_a}^2$,
\begin{equation}
 \hs{D\Gamma_P^*y}^2
 \le2\sum_{a,b}\Bigl(\hs{P_aY_aP_b}^2+\hs{P_aY_bP_b}^2\Bigr)
 \le4\sum_a\hs{Y_a}^2=4d\norm y_2^2 ,
 \label{eq:s-adjoint-upper}
\end{equation}
so $\norm{D\Gamma_P}=\norm{D\Gamma_P^*}\le2\sqrt d$. For a second
decomposition $P'$, the bound $\hs{[Y,X]}\le2\norm Y\hs X$,
Cauchy--Schwarz, and $\sum_a\norm{Y_a}^2\le\sum_a\hs{Y_a}^2=d\norm y_2^2$
give
\begin{equation}
 \hs{(D\Gamma_{P'}^*-D\Gamma_P^*)y}
 \le2\sqrt d\,\norm y_2\Bigl(\sum_a\hs{P'_a-P_a}^2\Bigr)^{1/2}.
 \label{eq:s-lipschitz}
\end{equation}
If $P_a=U\bar P_aU^\dagger$ and $P'_a=V\bar P_aV^\dagger$, then
$P'_a-P_a=(V-U)\bar P_aV^\dagger+U\bar P_a(V-U)^\dagger$, so
$\sum_a\hs{P'_a-P_a}^2\le4\sum_a\hs{(V-U)\bar P_a}^2=4\hs{U-V}^2$, and
the right side of Eq.~\eqref{eq:s-lipschitz} is at most
$4\sqrt d\,\norm y_2\hs{U-V}$.

\begin{proof}[Proof of Lemma~\ref{lem:adjoint}]
Write $P_a=U\bar P_aU^\dagger$, recall $r_a\ge d/(2q)$, and for a unit
vector $y\in\mathcal Y$ let $f_y(U)=\hs{D\Gamma_P^*y}$. Then
$\sigma(P)$ is the minimum of $f_y$ over all unit vectors $y$.

\emph{Step 1. Mean square for one direction.} For $a\ne b$ the blocks
$a$ and $b$ consist of different basis vectors, so $\hs{P_aZP_b}^2$ is
the sum of $|\bra{Ux}Z\ket{Ux'}|^2$ over the $r_ar_b$ pairs of basis
vectors $\ket x$ of block $a$ and $\ket{x'}$ of block $b$. The
twirl~\eqref{eq:s-weingarten} with $X=\ket{x'}\bra x$, which has
$\Tr X=0$ and $\hs X=1$, gives
$\E|\bra{Ux}Z\ket{Ux'}|^2=\hs Z^2/(d^2-1)$ for traceless $Z$. Hence
\begin{equation}
 \E\,\hs{P_aZP_b}^2=\frac{r_ar_b}{d^2-1}\hs{Z}^2 .
 \label{eq:s-offdiag-moment}
\end{equation}
Apply this to $Z=Y_a-Y_b$ in Eq.~\eqref{eq:s-adjoint-norm}, with
$\hs{Y_a-Y_b}^2=d\norm{y_a-y_b}_2^2$. Since
$r_ar_bd/(d^2-1)\ge r_ar_b/d\ge d/(4q^2)$ and, by the constraint
$\sum_ay_a=0$,
$\sum_{a\ne b}\norm{y_a-y_b}_2^2
 =2q\sum_a\norm{y_a}_2^2-2\bigl\|\sum_ay_a\bigr\|_2^2=2q$,
\begin{equation}
 \E f_y^2\ \ge\ \frac{d}{4q^2}\cdot2q=\frac{d}{2q}.
 \label{eq:s-adjoint-mean}
\end{equation}

\emph{Step 2. Concentration for one direction.} The function $f_y$
depends on $U$ only through the projectors $P_a$, so it is invariant
under a global phase, and by Eq.~\eqref{eq:s-lipschitz} it is
$4\sqrt d$-Lipschitz. Equation~\eqref{eq:s-concentration} thus gives
$\Pr\{|f_y-\E f_y|>t\}\le2e^{-ct^2/16}$. The fluctuations are
therefore of order one,
$\operatorname{Var}f_y=\int_0^\infty2t\Pr\{|f_y-\E f_y|>t\}\,dt\le32/c$,
and $(\E f_y)^2=\E f_y^2-\operatorname{Var}f_y\ge d/(4q)$ once
$d\ge128q/c$. Since $\E f_y\ge\frac12\sqrt{d/q}$, the choice
$t=\frac14\sqrt{d/q}$ gives
\begin{equation}
 \Pr\Bigl\{f_y<\tfrac14\sqrt{d/q}\Bigr\}\le2e^{-cd/(256q)} .
 \label{eq:s-median}
\end{equation}

\emph{Step 3. All directions at once.} By
Eq.~\eqref{eq:s-adjoint-upper}, $|f_y-f_{y'}|\le2\sqrt d\,\norm{y-y'}_2$
for unit vectors $y,y'\in\mathcal Y$. Let $\mathcal M$ be a
$\delta$-net of the unit sphere of $\mathcal Y$ with
$\delta=1/(16\sqrt q)$. Since $\dim\mathcal Y=s(q-1)$, it can be
chosen with $|\mathcal M|\le(48\sqrt q)^{sq}$ points. If
$f_y\ge\frac14\sqrt{d/q}$ at every net point, then every unit $y$
satisfies $f_y\ge\frac14\sqrt{d/q}-2\sqrt d\,\delta=\frac18\sqrt{d/q}$,
that is, $\sigma(P)\ge\frac18\sqrt{d/q}$. A union bound over the net
gives
\begin{equation}
 \Pr\Bigl\{\sigma(P)<\tfrac18\sqrt{d/q}\Bigr\}
 \le2(48\sqrt q)^{sq}e^{-cd/(256q)}\le\frac18
 \label{eq:s-inf-bound}
\end{equation}
whenever $d\ge Csq^2\log(2q)$, which also covers $d\ge128q/c$.
\end{proof}

\begin{proof}[Proof of Lemma~\ref{lem:certificate}]
\emph{Step 1. Response along the correcting rotations.} For
$y\in\mathcal Y$ let $P_a(y)=e^{Ty}P_ae^{-Ty}$ and
$F(y)=\Gamma(P(y))\in\mathcal Y$, so that $F(0)=\Gamma(P)$. Writing
$K_t=Ty_t$, the iteration of Sec.~\ref{sec:newton} reads
$y_{t+1}=y_t-F(y_t)$ with $y_0=0$. For a direction $y'\in\mathcal Y$, differentiating the
rotation gives
\begin{equation}
 DF(y)y'=D\Gamma_{P(y)}(\Xi),\qquad
 \Xi=\int_0^1e^{uTy}(Ty')e^{-uTy}\,du ,
 \label{eq:s-rotated-derivative}
\end{equation}
so $\Xi$ is the generator $Ty'$ averaged along the rotation. Since
$D\Gamma_P(Ty')=y'$,
$DF(y)y'-y'=(D\Gamma_{P(y)}-D\Gamma_P)(\Xi)+D\Gamma_P(\Xi-Ty')$.
With $K=Ty$, each difference
$P_a(y)-P_a=\int_0^1e^{uK}[K,P_a]e^{-uK}\,du$ has Hilbert--Schmidt
norm at most $\hs{[K,P_a]}$, and
$\sum_a\hs{[K,P_a]}^2=2\hs K^2-2\sum_a\hs{P_aKP_a}^2\le2\hs K^2$.
Equation~\eqref{eq:s-lipschitz} therefore bounds the first term by
$2\sqrt{2d}\,\hs K\hs{\Xi}$. By Eq.~\eqref{eq:s-adjoint-upper} the
second term is at most $2\sqrt d\,\hs{\Xi-Ty'}$, and
$\hs{\Xi-Ty'}\le\norm K\hs{Ty'}$ because
$\hs{e^{uK}(Ty')e^{-uK}-Ty'}\le2u\norm K\hs{Ty'}$. With
$\hs{\Xi}\le\hs{Ty'}\le\norm{y'}_2/\sigma$ and
$\hs K\le\norm y_2/\sigma$,
\begin{equation}
 \norm{DF(y)-\Id}\le(2\sqrt2+2)\frac{\sqrt d\,\norm y_2}{\sigma^2}
 \le\frac{5\sqrt d\,\norm y_2}{\sigma^2}.
 \label{eq:s-derivative-bound}
\end{equation}

\emph{Step 2. Rotating against the residual is a contraction.} On the
ball $\norm y_2\le2\beta$ of $\mathcal Y$, Eq.~\eqref{eq:s-derivative-bound}
and the first condition in Eq.~\eqref{eq:s-certificate-conditions}
give $\norm{DF(y)-\Id}\le10\sqrt d\,\beta/\sigma^2\le\frac12$. The map
$y\mapsto y-F(y)$ therefore shrinks all distances in the ball by at
least a factor of two. It also maps the ball into itself, because
$\norm{y-F(y)}_2\le\norm{F(0)}_2+\frac12\norm y_2\le2\beta$. By the
contraction principle, the iterates $y_t$ converge to a point $\hat y$
of the ball with $F(\hat y)=0$.

\emph{Step 3. The corrected measurement.} Put $K=T\hat y$ and
$\tilde P_a=P_a(\hat y)$. Then $\Gamma(\tilde P)=F(\hat y)=0$, which means
$\Tr(\tilde P_aA_\ell)=0$ for all $a,\ell$, and the rotation preserves
ranks, orthogonality, and completeness. Moreover
$\norm K\le\hs K\le2\beta/\sigma$, and for a contraction $A$,
\begin{equation}
 \norm{\tilde P_aA\tilde P_a}=\norm{P_a\,e^{-K}Ae^{K}P_a}
 \le\norm{P_aAP_a}+2\norm{e^K-\Id}
 \le b_{\max}+\frac{4\beta}{\sigma}\le\epsilon ,
 \label{eq:s-perturbation}
\end{equation}
where $\norm{e^K-\Id}\le\norm K$ was used and the last step is the
second condition in Eq.~\eqref{eq:s-certificate-conditions}.
\end{proof}

\begin{proof}[Proof of Proposition~\ref{prop:sectors}]
\emph{Step 1. The schedule.} Put $\theta=\log(1/\eta)$ and
$\theta_0=\log5$, and let $L\ge1$ be the largest integer with
$2^{L-1}\theta_0\le\theta$. Define
\begin{equation}
 \epsilon_j=\eta^{2^{-j}}\quad(1\le j<L),\qquad
 \epsilon_L=\eta^{2^{1-L}},\qquad
 q_j=\lceil C_0\epsilon_j^{-2}\rceil,\qquad
 Q_j=\prod_{h\le j}q_h,\quad Q_0=1 .
 \label{eq:s-hierarchy-parameters}
\end{equation}
Every level $j<L$ supplies half of the exponent of $\eta$ that is
still missing, and the last level supplies the rest, so
$\prod_j\epsilon_j=\eta$ because $\sum_{j<L}2^{-j}+2^{1-L}=1$. All
exponents are at least $2^{1-L}$, and $2^{1-L}\theta\ge\theta_0$ by
the choice of $L$, so every $\epsilon_j\le1/5<1/4$, as
Lemma~\ref{lem:onestep} requires. Maximality of $L$ gives
$2^{1-L}\theta<2\theta_0$, so the last factor obeys
$\epsilon_L>1/25$. Finally, $q_j\le C_*\epsilon_j^{-2}$ with
$C_*=C_0+1$.

\emph{Step 2. Induction over levels.} Suppose that after level $j-1$
the space is split into $Q_{j-1}$ sectors, with mutually orthogonal
projections $\Pi_v$ summing to $\Id_d$, isometries
$J_v:\C^{d_v}\to\C^d$ onto their ranges, dimensions
$\lfloor d/Q_{j-1}\rfloor\le d_v\le\lceil d/Q_{j-1}\rceil$, and
\begin{equation}
 \Tr\bigl(J_v^\dagger A_\ell J_v\bigr)=0,
 \qquad
 \norm{J_v^\dagger A_\ell J_v}\le\rho_{j-1}
 =\prod_{h<j}\epsilon_h
 \qquad\text{for all }\ell,v .
 \label{eq:s-inductive-hypothesis}
\end{equation}
This holds at level $0$, with the single sector $\C^d$ and
$\rho_0=1$. In the sector $v$, the rescaled operators
\begin{equation}
 A_\ell^{(v)}=\rho_{j-1}^{-1}J_v^\dagger A_\ell J_v
 \label{eq:s-rescaled}
\end{equation}
are traceless contractions on $\C^{d_v}$. Their Hermitian parts are
the compressions of $\RePart A_\ell$ and $\ImPart A_\ell$, so their
real span is the image of the span in Eq.~\eqref{eq:s-spandim} under
the real linear map $X\mapsto\rho_{j-1}^{-1}J_v^\dagger XJ_v$ and has
dimension at most $s$. Lemma~\ref{lem:onestep} with
$(\epsilon_j,q_j)$ therefore applies in every sector as soon as
\begin{equation}
 \left\lfloor\frac d{Q_{j-1}}\right\rfloor
 \ \ge\ C_1\!\left[sq_j^2\log(2q_j)
 +\epsilon_j^{-2}\log(2mq_j)\right].
 \label{eq:s-level-condition}
\end{equation}
It splits each sector into $q_j$ balanced subsectors with zero block
traces and compressed norms at most $\rho_{j-1}\epsilon_j=\rho_j$.
Their dimensions lie between $\lfloor d_v/q_j\rfloor$ and
$\lceil d_v/q_j\rceil$, hence between $\lfloor d/Q_j\rfloor$ and
$\lceil d/Q_j\rceil$, because nested integer divisions compose,
$\lfloor\lfloor d/Q_{j-1}\rfloor/q_j\rfloor=\lfloor d/Q_j\rfloor$ and
$\lceil\lceil d/Q_{j-1}\rceil/q_j\rceil=\lceil d/Q_j\rceil$. This is
Eq.~\eqref{eq:s-inductive-hypothesis} at level $j$. After level $L$,
the $Q_L$ final sectors form a single projective decomposition
$\{\Pi_w\}$ with
\begin{equation}
 \Tr(\Pi_wA_\ell)=0,
 \qquad
 \norm{\Pi_wA_\ell\Pi_w}\le\prod_{j=1}^L\epsilon_j=\eta ,
 \label{eq:s-leaf-conclusion}
\end{equation}
and transposing its projectors gives Alice's measurement.

\emph{Step 3. Dimension cost at each level.} It remains to show that
Eq.~\eqref{eq:s-final-dimension} implies
Eq.~\eqref{eq:s-level-condition} at every level. At level $j$ each of
the $Q_{j-1}$ sectors already present pays the refinement cost
$sq_j^2$. Since $Q_{j-1}\le C_*^{j-1}(\epsilon_1\cdots\epsilon_{j-1})^{-2}$
and $q_j^2\le C_*^2\epsilon_j^{-4}$, the schedule gives
\begin{equation}
 Q_{j-1}q_j^2\le C_*^{j+1}\,\eta^{-2+2^{2-j}}\,\eta^{-2^{2-j}}
 =C_*^{j+1}\eta^{-2}\quad(j<L),
 \qquad
 Q_{L-1}q_L^2\le C_*^{L+1}\eta^{-2}\epsilon_L^{-2}
 \le625\,C_*^{L+1}\eta^{-2}.
 \label{eq:s-costs-raw}
\end{equation}
Similarly,
$Q_{j-1}\epsilon_j^{-2}\le C_*^{j-1}(\epsilon_1\cdots\epsilon_j)^{-2}
\le C_*^L\eta^{-2}$ at every level.

\emph{Step 4. Collecting the logarithms.} The number of levels obeys
$L\le1+\log_2(\theta/\theta_0)$, so
$C_*^L\le C(1+\theta)^{\log_2C_*}$. Moreover
$\log(2q_j)\le C(1+\theta)$, $\log(2mq_j)\le C\log(2m/\eta)$, and
$1+\theta\le2\log(2m/\eta)$ for $\eta\le1/8$. Multiplying
Eq.~\eqref{eq:s-level-condition} by $Q_{j-1}$, and using
$\lfloor x\rfloor\ge x/2$ for $x\ge1$, Eq.~\eqref{eq:s-costs-raw}
shows that all level conditions follow from
Eq.~\eqref{eq:s-final-dimension} with $\nu=1+\log_2C_*$. The number of
outcomes is then
$Q=Q_L\le C_*^L\eta^{-2}\le C\eta^{-2}[\log(2m/\eta)]^\nu$, which
completes the proof.
\end{proof}

\section{Consequences and extensions}
\label{sec:extensions}

The results below build on Theorem~1. The extension to
nonmaximally entangled states also reuses Proposition~\ref{prop:sectors}
and Bob's measurement of Sec.~\ref{sec:bob}.

\subsection{Measurement hierarchy}
\label{sec:s-hierarchy-thresholds}

For $d\ge2$, let $m_{\mathsf M}(d)$ be the smallest size of an
orthogonal MES set in $\C^d\otimes\C^d$ that is not perfectly
distinguishable by class $\mathsf M$. Every smaller set is
therefore distinguishable. The subscript $\rightarrow$ denotes
one-way LOCC from Alice to Bob, while $\leftrightarrow$ denotes
two-way LOCC with any finite number of communication rounds.
Inverting Eq.~\eqref{eq:s-main-bound} gives
\begin{equation}
 m_{\leftrightarrow}(d)
 =\Omega\!\left[\frac{d^{1/4}}{(\log d)^{\nu/4}}\right],
 \label{eq:s-cardinality}
\end{equation}
using only one feedback message. For comparison, separable (SEP)
measurements have effects that are sums of positive product
operators, and positive partial transpose (PPT) measurements have
effects that remain positive under partial transpose on Bob's side.
The inclusions one-way LOCC $\subseteq$ two-way LOCC $\subseteq$ SEP
$\subseteq$ PPT imply
\[
 m_{\rightarrow}(d)\le m_{\leftrightarrow}(d)
 \le m_{\rm SEP}(d)\le m_{\rm PPT}(d).
\]
Thus Eq.~\eqref{eq:s-cardinality} also applies to SEP and PPT. For
PPT it is superseded by $m_{\rm PPT}(d)\ge\lfloor d/2\rfloor+2$
from Ref.~\cite{Nathanson2013}. Table~\ref{tab:s-hierarchy} includes
additional bounds for particular dimensions.

\begin{table}[htbp]
\caption{Detailed thresholds for perfect discrimination of orthogonal
MES sets using one copy and no additional shared entanglement.
A lower bound guarantees that every smaller set is distinguishable;
an upper bound is witnessed by an indistinguishable set of no more
than that size. The arrows denote the communication classes defined
above, and the two-way lower bound uses one feedback message.
The notation $d^{1/4-o(1)}$ suppresses the logarithmic factor in
Eq.~\eqref{eq:s-cardinality}. $(*)$ marks this work.}
\label{tab:s-hierarchy}
\centering
\small
\setlength{\tabcolsep}{6pt}
\renewcommand{\arraystretch}{1.2}
\begin{tabular}{llll}
\toprule
Class $\mathsf M$ & Lower bound on $m_{\mathsf M}(d)$ & Upper bound on $m_{\mathsf M}(d)$ & Open question \\
\midrule
One-way LOCC ($\rightarrow$) & $3$ \cite{Walgate2000} &
\begin{tabular}[t]{@{}l@{}}
$4$ (all $d\ge4$) \cite{Wang2016}\\
$3$ (even $d$ or $d\equiv2\pmod 3$) \cite{Nathanson2013}
\end{tabular}
& $m_{\rightarrow}(d)=3$ for all $d\geq4$? \\
\addlinespace[3pt]
Two-way LOCC ($\leftrightarrow$) & $d^{1/4-o(1)}$ $(*)$ &
\begin{tabular}[t]{@{}l@{}}
$d$ (all $d\ge4$) \cite{Yu2012,YuOh2015}\\
$d-1$ (even $d\ge6$) \cite{YuOh2015}\\
$3d/4+1$ ($d\equiv0\pmod 4$) \cite{YuOh2015}
\end{tabular}
& $m_{\leftrightarrow}(d)=\Theta(d)$? \\
\addlinespace[3pt]
Separable (SEP) & $d^{1/4-o(1)}$ $(*)$ &
\begin{tabular}[t]{@{}l@{}}
$d-1$ (even $d\ge6$) \cite{YuOh2015}\\
$3d/4+1$ ($d\equiv0\pmod 4$) \cite{YuOh2015}
\end{tabular}
& $m_{\rm SEP}(d)=\Theta(d)$? \\
\addlinespace[3pt]
PPT & $\lfloor d/2\rfloor+2$ \cite{Nathanson2013} &
\begin{tabular}[t]{@{}l@{}}
$\lfloor7d/8\rfloor+1$ ($d=2^n$, $n\ge4$) \cite{CosentinoRusso2014}\\
$4$ ($d=4$) \cite{Yu2012}
\end{tabular}
& asymptotic ratio $m_{\rm PPT}(d)/d$? \\
\bottomrule
\end{tabular}
\end{table}

\subsection{Multiple copies}

\begin{restatedcorollary}[A universal copy bound]
There is a universal integer $N_0$ such that every orthogonal MES
set, in every local dimension $d\ge2$, is perfectly distinguishable
from $N_0$ copies by an $A\to B\to A$ protocol. For all sufficiently
large $d$, nine copies suffice.
\end{restatedcorollary}

For $d\ge2$, allow each party joint local operations on its $N$ copies.
After grouping Alice's systems and Bob's systems,
$\ket{\Phi_d}^{\otimes N}$ becomes the maximally entangled state
$\ket{\Phi_{d^N}}$ of rank $d^N$, so
$\ket{\Psi_i}^{\otimes N}=(\Id\otimes U_i^{\otimes N})\ket{\Phi_{d^N}}$.
The labels remain orthogonal because
$\Tr[(U_i^\dagger U_j)^{\otimes N}]=[\Tr(U_i^\dagger U_j)]^N=0$.
Theorem~1 therefore applies for $k\ge3$ whenever
\begin{equation}
 N\ge\max\!\left\{1,
 \left\lceil\log_d\!\bigl(Ck^4[\log(2k)]^\nu\bigr)\right\rceil\right\}.
 \label{eq:s-multicopy}
\end{equation}
For $k\le2$, one copy suffices~\cite{Walgate2000}.
Since $k\le d^2$, the dimension threshold is at most
$Cd^8[\log(2d^2)]^\nu\le C'd^9$ for a universal $C'\ge1$.
Thus $N_0=9+\lceil\log_2C'\rceil$ suffices for every $d\ge2$.
Moreover $C[\log(2d^2)]^\nu/d\to0$, so nine copies suffice for all
sufficiently large $d$, which proves Corollary~1.

For general orthogonal pure states, $k-1$ copies
suffice~\cite{Walgate2000,Bandyopadhyay2011}. This is tight for some
bases of two qubits when each copy is measured separately and later
measurements depend on earlier outcomes~\cite{Banik2021}. That
restriction is absent here. Two copies suffice for subsets of the
canonical maximally entangled basis~\cite{Ghosh2004}; Corollary~1
requires no such basis structure.

\subsection{Nonmaximally entangled states}
\label{sec:bounded-schmidt}

Let $V_i:\C^d\to\C^D$, $D\ge d$, satisfy
$\Tr(V_i^\dagger V_j)=d\delta_{ij}$ and
$\lambda_-\Id\preceq V_i^\dagger V_i\preceq\lambda_+\Id$, with
$0<\lambda_-\le1\le\lambda_+$ and $\kappa=\lambda_+/\lambda_-$.
The states $(\Id\otimes V_i)\ket{\Phi_d}$, written
$d^{-1/2}\operatorname{vec}(V_i)$ in the main text, are orthogonal,
and their squared Schmidt coefficients lie in
$[\lambda_-/d,\lambda_+/d]$. Thus $\kappa$ bounds the ratio of the
largest to the smallest eigenvalue of each reduced state on Alice's
side.
For $k\ge3$, Proposition~\ref{prop:sectors} applies to
$A_{ij}=V_i^\dagger V_j/\lambda_+$ with
$\eta'=1/[4\kappa(k-1)]$ and $s\le k(k-1)$ whenever
\begin{equation}
 d\ge C\kappa^2k^4[\log(2\kappa k)]^\nu.
 \label{eq:s-bounded-schmidt-threshold}
\end{equation}
This is the bound stated in the main text.
On a sector of rank $r$ with isometry $J:\C^r\to\C^d$, $JJ^\dagger=P_a$,
let $W_i=V_iJ$,
$C_{ij}=W_i^\dagger W_j$ and $t_i=\Tr C_{ii}>0$. Then
\begin{equation}
 C_{ii}\succeq\lambda_-\Id_r,\qquad \Tr C_{ij}=0,\qquad
 \norm{C_{ij}}\le\frac{\lambda_-}{4(k-1)}\quad(i\ne j).
 \label{eq:s-bounded-schmidt-data}
\end{equation}
Alice's outcome now has probability $t_i/d$, which may depend on
the label, and leaves the normalized state
$\sqrt{r/t_i}\,(\Id\otimes W_i)\ket{\Phi_r}$.
The proof of Proposition~\ref{lem:gram} gives
$G\succeq\tfrac34\lambda_-\Id$ and
$R_i=C_{ii}-\sum_{j\ne i}D_{ij}\succeq\tfrac34\lambda_-\Id_r$.
Hence the POVM and support identities in
Eqs.~\eqref{eq:s-povm}--\eqref{eq:s-support} still apply.
For an effect $B$ and a pair outcome implemented with
$B_{ij}^{1/2}$, respectively,
\[
 p(B|i)=\frac{\Tr(W_i^\dagger BW_i)}{t_i},\qquad
 \frac{\Tr(W_i^\dagger B_{ij}W_j)}{\sqrt{t_it_j}}
 =\frac{\Tr C_{ij}}{\sqrt{t_it_j}}=0.
\]
Thus the same outcomes exclude unwanted labels and preserve purity
and pair orthogonality. Bob starts the Walgate protocol and sends his
outcome to Alice, who completes the discrimination. For $\kappa=1$,
normalization forces $V_i^\dagger V_i=\Id_d$, recovering MESs of
Schmidt rank $d$ in $\C^d\otimes\C^D$. For $k\le2$, no dimension or
spectral condition is needed~\cite{Walgate2000}. This extension assumes
exactly orthogonal pure states, not tolerance to experimental errors.

\subsection{Decodable message length}

For finite $m_{\mathsf M}(d)$, define $b_{\mathsf M}(d)$ as the largest
nonnegative integer $b$ for which every codebook of at most $2^b$
orthogonal MESs is perfectly distinguishable by $\mathsf M$. Then
\begin{equation}
 b_{\mathsf M}(d)=\left\lfloor\log_2[m_{\mathsf M}(d)-1]\right\rfloor,
 \label{eq:s-bits-definition}
\end{equation}
since every codebook of at most $2^b$ states is distinguishable
exactly when
$2^b<m_{\mathsf M}(d)$.

\begin{proposition}[Decodable bits]
\label{prop:bits}
For every $d\ge4$, $b_{\rightarrow}(d)=1$. For $d=2^n\ge4$,
\begin{equation}
 \frac n4-O(\log n)\le b_{\leftrightarrow}(d)\le n-1.
 \label{eq:s-bits-bounds}
\end{equation}
\end{proposition}
\begin{proof}
Distinguishability of every pair~\cite{Walgate2000} and one-way
counterexamples with four states~\cite{Wang2016} give
$3\le m_{\rightarrow}(d)\le4$, hence $b_{\rightarrow}(d)=1$.
Equation~\eqref{eq:s-cardinality} gives
$m_{\leftrightarrow}(d)\ge cd^{1/4}/(\log d)^{\nu/4}$ and therefore
$b_{\leftrightarrow}(d)\ge n/4-O(\log n)$.
For $d\ge4$, the existence of sets of $d$ MESs that LOCC cannot
distinguish~\cite{Yu2012,YuOh2015} gives
$b_{\leftrightarrow}(d)\le\lfloor\log_2(d-1)\rfloor=n-1$.
\end{proof}
These message lengths hold for \emph{every} codebook. They are not
an optimized capacity, nor the communication cost of the decoding
protocol.

\FloatBarrier

\bibliography{references}